\documentclass[apj,twocolumn]{openjournal}
\usepackage{amsmath}
\usepackage{amssymb}
\usepackage{booktabs}
\usepackage{multirow}
\usepackage{color}
\usepackage{soul}
\usepackage[dvipsnames]{xcolor}
\usepackage[breaklinks,colorlinks,citecolor=blue,urlcolor=blue]{hyperref}
\usepackage{listings}
\usepackage{orcidlink}

\renewcommand{\arraystretch}{1.2}
\begin{document}
\makeatletter
\let\frontmatter@title@above=\relax
\makeatother

\newcommand\lsim{\mathrel{\rlap{\lower4pt\hbox{\hskip1pt$\sim$}}
\raise1pt\hbox{$<$}}}
\newcommand\gsim{\mathrel{\rlap{\lower4pt\hbox{\hskip1pt$\sim$}}
\raise1pt\hbox{$>$}}}
\newcommand{\CS}[1]{{\color{red} CS: #1}}

\title{How precisely can we measure the ages of subgiant and giant stars?}

\shorttitle{Precision of subgiant and giant star ages}
\shortauthors{Shariat, El-Badry, \& Bhattacharjee}

\author{\vspace{-30pt}Cheyanne Shariat\,\orcidlink{0000-0003-1247-9349}$^{1}$}
\author{Kareem El-Badry\,\orcidlink{0000-0002-6871-1752}$^{1}$}
\author{Soumyadeep Bhattacharjee\,\orcidlink{0000-0003-2071-2956}$^{1}$}

\affiliation{$^1$Department of Astronomy, California Institute of Technology, 1200 East California Boulevard, Pasadena, CA 91125, USA}
\email{Corresponding author: cshariat@caltech.edu}


\begin{abstract}
Precise stellar ages are fundamental to Galactic archaeology. However, obtaining reliable age estimates and uncertainties for field stars has been a long-standing challenge. We test the fidelity of ages from recent catalogs of giants and subgiants using wide binaries, whose components formed at the same time and thus should have consistent inferred ages. We find that subgiant ages based on spectroscopic metallicities from \citet{XiangRix2022} are generally consistent within their reported uncertainties, implying that fractional uncertainties of $5-10\%$ are realistically achievable. In contrast, we find that published photometric subgiant ages underestimate true uncertainties by factors of $2-3$. Spectroscopic age estimates for red giant and red clump stars also show reliable uncertainties, but are generally less precise ($25-30\%$). These results demonstrate that accurate chemical abundance measurements are essential for precise subgiant ages and establish wide binaries as a powerful, model-independent benchmark for calibrating stellar age measurements in the era of large spectroscopic surveys.
\end{abstract}

\maketitle

\section{Introduction}\label{sec:intro}

Reconstructing the Milky Way’s formation and chemical enrichment history requires precise ages for large numbers of stars across the Galaxy \citep[e.g.,][]{Twarog80,Soderblom2010,Casagrande16,Sanders18,Deason24}. Modern surveys now routinely estimate ages for millions of field stars by combining astrometry, multi-band photometry, spectroscopy, and, where available, asteroseismology \citep[e.g.,][]{Xiang17,Queiroz18,Sanders18,Helmi20,XiangRix2022,Nataf2024,Wang2023,Wang25}. Among different evolutionary stages, subgiants are especially powerful chronometers: their luminosities depend sensitively on core mass, which scales with stellar mass and thus main-sequence lifetime. With sufficiently precise parallaxes and metallicity measurements, isochrone fitting can 
achieve relative age precisions at the percent level for subgiant populations \citep[e.g.,][]{Nataf2024}.

For Galactic archaeology, achieving stellar ages with $\lesssim10\%$ precision over a wide age range ($\sim1-14$~Gyr) is critical for reconstructing the Galaxy’s star formation history and quantifying its chemical evolution \citep[e.g.,][]{Xiang17,Bonaca20,Helmi20}. However, estimating accurate age uncertainties remains challenging. Even small differences in how ages are estimated can lead to different conclusions about the timeline of the Galaxy's disk and halo assembly \citep[e.g.,][]{Conroy22, XiangRix2022}. 
Many internal validation tests rely on stellar evolution models, often the same ones used to derive the ages, which limits their diagnostic power and fails to probe systematic errors fully. Empirical checks using open and globular clusters provide an important comparison, but such clusters span a narrow range of ages and metallicities. An independent benchmark that samples a broader range of stellar populations is therefore essential.

Wide binaries provide such a benchmark. They are distributed across the Galactic disk and halo, and effectively sample the entire range of stellar ages and metallicities found in the Milky Way. 
Their components share a common chemical composition \citep[e.g.,][]{Hawkins20} and are coeval \citep[i.e., effectively the same age;][]{Makarov08,Kraus09}. 
This makes them uniquely valuable for testing whether reported age uncertainties are realistic: discrepancies between the two components, normalized by the quoted errors, directly reveal whether catalog error bars are under- or overestimated. The coeval and co-chemical nature of wide binaries has enabled progress across many areas of stellar astrophysics, including the calibration of gyrochronology \citep[e.g.,][]{Barnes07,Mamajek08,Deacon16,GodoyRivera18,Otani22,Gruner23,SilvaBeyer23,LaresMartiz24}, M-dwarf metallicity estimates \citep[e.g.,][]{Bonfils05,Lepine07,Johnson09,RojasAyala10,Mann13,Montes18}, stellar activity-age relations \citep[e.g.,][]{Garces11,Chaname12}, consistency tests for chemical tagging \citep[e.g.,][]{Andrews19,Hawkins20,EB21_widebin}, the age--metallicity relation \citep[][]{RebassaMansergas16}, and the initial--final mass relation for white dwarfs \citep[][]{Zhao12,Andrews15,Barrientos21,Hollands24}. For a recent review discussing wide binaries with {\it Gaia}, see \citet{EB24_review}.

In this work, we use wide binaries in which both components are evolved stars with independently inferred ages to test the accuracy of reported age uncertainties in a model-independent way. The remainder of this paper is organized as follows. In Section \ref{sec:methods}, we describe the sample selection methodology. Section \ref{sec:results} presents our main results, and Section \ref{sec:discussion} provides a discussion of our results. Finally, Section \ref{sec:conclusions} summarizes our conclusions and considers future directions.

\section{Data and Sample Selection}\label{sec:methods}
\subsection{Parent catalogs}\label{subsec:parents}

Catalogs providing ages for a large sample of evolved stars have recently been constructed using both photometric and spectroscopic data. The three catalogs that we focus on are those of \citet{XiangRix2022}, \citet{Nataf2024}, and \citet{Wang2023}. We briefly summarize each below. 

\citet{XiangRix2022} constructed a subgiant catalog by combining {\it Gaia} astrometry with LAMOST DR7 spectroscopy. They first select subgiants using an HR diagram cut.
Stellar atmospheric parameters ($T_{\rm eff}$, [Fe/H], [$\alpha$/Fe]) are derived for all spectra with SNR $>20$, restricting to stars with $T_{\rm eff} \leq 6800$~K, where the spectral fits are more robust. 
They emphasize that accurate abundances are crucial for dating subgiants, showing that even modest [$\alpha$/Fe] offsets (e.g., $0.20$~dex) can shift isochrone ages by $1-2$~Gyr \citep[their Extended Data Fig.~4;][]{XiangRix2022}, broadly consistent with the expectation that $\alpha$-enhancement modifies the effective metal content of stellar models \citep{Salaris1993}. Stars with $M_K < 0.5$~mag were excluded to avoid contamination from He-burning horizontal branch stars. Also, objects whose spectro-photometric and {\it Gaia} parallax distances disagreed by more than $2\sigma$ were removed, as were unresolved binaries flagged by {\it Gaia} diagnostics. The final catalog thus minimizes contamination from misclassified evolutionary states and unresolved companions. Ages were inferred by fitting the derived parameters ($T_{\rm eff}$, $M_K$, [Fe/H], [$\alpha$/Fe]) together with {\it Gaia}+2MASS photometry to a grid of Yonsei-Yale (Y$^2$) isochrones \citep{Yi01,Demarque04}. 

\citet{Nataf2024} developed a photometric pipeline to infer subgiant ages from \emph{Gaia} DR3 distances and UV-IR photometry. Subgiants were selected via a cut in the color-magnitude diagram (CMD), but since their locus overlaps with both the main-sequence turn off (MSTO) and base of the red giant branch (RGB), the cut is strongly age- and metallicity-dependent. To mitigate this, they split the sample into two: a ``Primary Sample'' targeting $-0.5\lesssim{\rm [Fe/H]}\lesssim0.5$, where subgiants are cleanly separated from metal-rich MSTO stars, and a metal-poor ``annex'' selected using GALEX NUV or SDSS/Skymapper $u$-band photometry. Additional astrometric and photometric quality cuts were applied to enhance purity and remove unresolved binaries. Ages were inferred using the {\tt isochrones} package \citep{Morton15}, which fits MIST isochrones to astrometry, photometry, and (implicitly constrained) metallicity. Unlike \citet{XiangRix2022}, metallicities here are not derived spectroscopically but are instead primarily inferred from UV photometry. Thus, any systematic mismatch between metallicity estimates propagates directly into the inferred ages. By comparing to \citet{XiangRix2022}, \citet{Nataf2024} find systematic offsets: their metallicities were typically higher by $0.19\pm0.10$~dex, and ages were lower, with an average ratio of $0.94\pm0.13$. They concluded that their UV photometry offers a comparable diagnostic to large spectroscopic surveys for subgiant age inference. Overall, they report a median age precision in their sample of $8-10\%$.

\citet{Wang2023} derived ages for $\sim10^6$ LAMOST DR8 RGB and red clump (RC) stars. Giants were selected by $T_{\rm eff}\leq5800$\,K and $\log g \leq 3.8$. 
To distinguish RGB from RC stars, they employed pseudo-asteroseismic diagnostics ($\Delta\nu$, $\Delta P$) inferred from spectra via a neural network trained on Kepler giant stars with asteroseismology. These quantities, when measured precisely, can effectively distinguish RGBs from RCs \citep[e.g.,][]{Bedding11,Stello13,Pinsonneault14,Vrard16,Elsworth17,Wu19}.
Because such labels exist for many \emph{Kepler} giants that are also observed by LAMOST, \citet{Wang2023} trained a neural network to predict $\Delta\nu$ and $\Delta P$ directly from LAMOST spectra (following \citealt{Ting18,Wu19}). This enabled RGB/RC classification and subsequent mass/age determination. For the training set (LAMOST+$\emph{Kepler}$), they fit PARSEC isochrones \citep{Bressan12} in a Bayesian framework, then applied the trained model to all DR8 giants. Typical uncertainties were $\sim27\%$ (age) and $6\%$ (mass) for RGBs and $\sim19\%$ (age) and $9\%$ (mass) for RCs. Comparisons against open clusters and other catalogs suggested that for spectra with SNR$>50$, age uncertainties are $\lesssim20\%$ for RGBs and $\lesssim25\%$ for RCs, with mass uncertainties below $10\%$.

\subsection{wide binary cross-match}\label{sec:wb}

We assess the reported age uncertainties using {\it Gaia} wide binaries. \citet{EB21_widebin} constructed a catalog of $\sim$1~million high-confidence wide binaries within $1$~kpc of the Sun using {\it Gaia} DR3 astrometry. We extend their sample to $5$~kpc using the exact same methodology, which preserves the original emphasis on purity while adding another $\sim500,000$ new wide binaries\footnote{The $5$~kpc wide binary catalog is available at \url{https://zenodo.org/records/17957444}.}. 
Because giant stars are intrinsically bright and thus maintain relatively low astrometric errors at distances above $1$~kpc, extending to $5$~kpc substantially boosts the number of usable systems for this study, despite parallax precision degrading with distance. The resulting catalog of $\sim$1.6~million pairs within $5$~kpc serves as the foundation for our analysis, from which we identify binaries where both components are evolved stars and have independently inferred ages from the parent catalogs.

We positionally cross-match each of the parent catalogs to the wide binary sample using a $1''$ tolerance. From these matches, we select systems in which {\it both} components of a wide binary are present in the same catalog and have reported ages. This yields $12$, $61$, and $76$ binaries for \citet{XiangRix2022}, \citet{Nataf2024}, and \citet{Wang2023}, respectively. We further subdivide the \citet{Nataf2024} sample to include only stars in their Primary Sample defined above ($N=16$), and the \citet{Wang2023} sample to those with spectra of SNR$>50$ ($N=21$).
Also, while not the focus of this study, requiring that only {\it one} wide binary component has an age estimate provides $3475$, $10226$, and $9184$ systems, which may be useful for other work. The full wide binary table is available in Appendix \ref{app:wbs}.
We note that, while the chance alignment probability ($\mathcal{R}$) for individual binaries is exceedingly low (Appendix \ref{app:wbs}), there is a $0.3\%$, $16\%$, and $40\%$ that at least one wide binary in the \citet{XiangRix2022}, \citet{Nataf2024}, and \citet{Wang2023} used in this analysis is a chance alignment.

\section{Results}\label{sec:results}

\begin{figure*}
\centering
\includegraphics[width=0.9\textwidth]{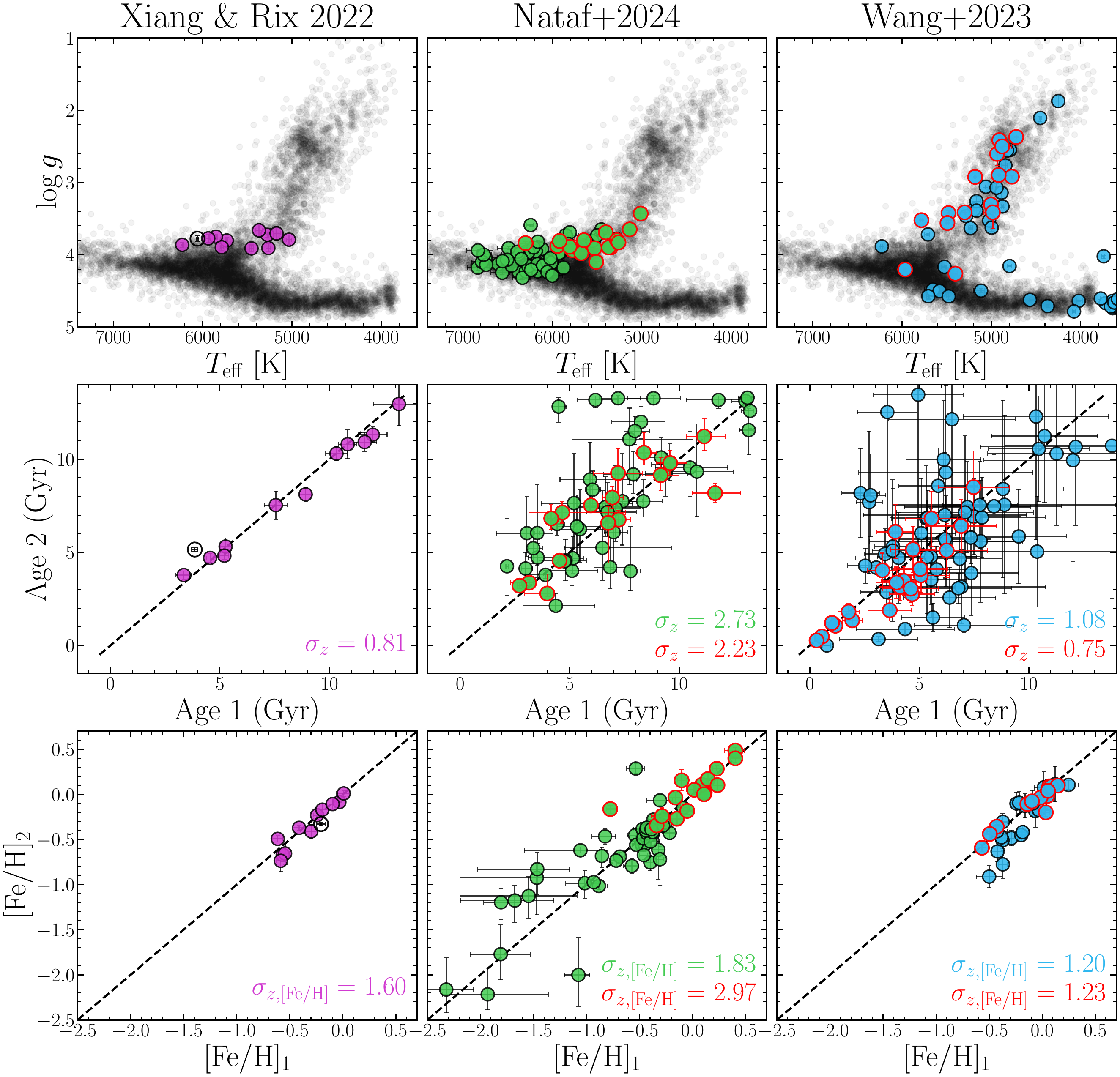}
\caption{
Calibrating subgiant and giant star age and metallicity uncertainties using wide binaries.
{\bf Top:} Kiel diagrams ($T_{\rm eff}$-$\log g$) for the stars in each sample: \citet{XiangRix2022} subgiants (purple; hollow points show the outlier), \citet{Nataf2024} subgiants (green, red outline for the Primary Sample), and \citet{Wang2023} LAMOST red giants (blue, red outline for SNR$>50$). The black points in the background show a randomly selected LAMOST sample with SNR$>50$ for reference.
{\bf Middle:} wide binary component age comparisons.
{\bf Bottom:} wide binary component metallicity comparisons for the same systems. Dashed lines show the $1:1$ line. The annotated $\sigma_z$ ($\sigma_{z,[{\rm Fe/H}]}$)  denote the standard deviations of the normalized age (metallicity) differences.
After removing one chemically discrepant outlier, the \citet{XiangRix2022} ages are consistent with their quoted uncertainties. The \citet{Nataf2024} sample shows substantially broader age and metallicity disagreement, while the \citet{Wang2023} sample remains broadly consistent, albeit with large age uncertainties.
}
\label{fig:RG_ages_accuracy}
\end{figure*}

To calibrate uncertainties, we compute the uncertainty-normalized age difference
\begin{equation}\label{eq:z}
z\;=\;\frac{{\rm Age}_1-{\rm Age}_2}{\sqrt{\sigma_{\rm Age,1}^2+\sigma_{\rm Age,2}^2}}\,
\end{equation}
for each of the samples, and determine its standard deviation, $\sigma_z$.
If quoted errors are realistic and independent, $z$ should be distributed as a Gaussian with unit variance $\sigma_z\approx1$ and mean $\langle z\rangle \approx0$. For \citet{Nataf2024}, who report asymmetric errors (e.g., 16th/84th percentiles), we symmetrize the uncertainty used in $z$ via the RMS of upper and lower errors, $\sigma=\sqrt{(\sigma_{\rm low}^2+\sigma_{\rm high}^2)/2}$. This approach avoids directional bias and provides a single representative $\sigma$ for computing $z$, while we retain the asymmetric error bars in visualizations\footnote{Omitting this symmetrization does not impact the broader results.}.

Figure~\ref{fig:RG_ages_accuracy} compares the component properties reported in the three catalogs. The top panels show the Kiel diagrams for each sample, the middle panels compare the ages of the primary (Age 1) and secondary (Age 2) components, and the bottom panels compare the corresponding metallicities, all with the published $1\sigma$ uncertainties shown. For the \citet{Nataf2024} catalog, we distinguish between their full sample and the ``Primary Sample'', which is restricted to $-0.5\lsim{\rm [Fe/H]}\lsim0.5$ (e.g., top middle of Figure~\ref{fig:RG_ages_accuracy}). For the \citet{Wang2023} sample, we additionally split their giants by spectral SNR, highlighting the SNR$>50$ subsample (for both stars), where they report typical age uncertainties of $\lsim25\%$. For each case, we compute and report $\sigma_z$, the standard deviation of the normalized age difference, and overplot the 1:1 relation for reference.

We find that the catalog of \citet{XiangRix2022} produces the most consistent subgiant ages. For the full sample, we find $\sigma_z=2.50$, which is mostly attributed to a single strong outlier with Age 1 = $3.847 \pm 0.141$~Gyr and Age 2 = $5.153 \pm 0.085$~Gyr. Removing this outlier reduces the spread to $\sigma_z=0.81$, fully consistent with the reported uncertainties. The outlier also shows a $3.25\sigma$ discrepant [Fe/H] between the two stars, likely explaining the divergent ages\footnote{We check that this source is indeed confidently a wide binary, with a chance alignment probability of $2.8\times10^{-4}$ \citep{EB21_widebin} at a physical separation of $\sim17,600$~au and distance $\sim800$~pc (Appendix \ref{app:wbs}).}. Despite the modest sample size ($N=11$), these results demonstrate that subgiant ages from \citet{XiangRix2022} are statistically consistent with their quoted uncertainties ($\sigma_z\approx1$).


For \citet{Wang2023}, the reported age errors are also consistent with the observed scatter: the full red-giant and red-clump sample yields $\sigma_z=1.08$, improving slightly to $\sigma_z=0.75$ for spectra with SNR$>50$. These results indicate that age precisions of $\sim25 - 30\%$ are realistically achievable for giants, though they remain less precise than the $\sim5 -10\%$ age uncertainties inferred for subgiants in \citet{XiangRix2022}.
There are 15 mixed RGB+RC systems in the \citet{Wang2023} sample used here. In these systems, the RC age depends on the mass-loss prescription adopted by \citet{Wang2023}. We therefore interpret the agreement in the \citet{Wang2023} sample as evidence that their full giant-star age pipeline works reasonably well. Such RGB+RC systems could in principle be used to constrain RGB mass loss, but we defer this to future work.

By contrast, the \citet{Nataf2024} catalog appears to underestimate age uncertainties. The full sample shows substantial excess scatter, and even in the Primary Sample, the dispersion remains well above unity ($\sigma_z\approx2.2$). This suggests that residual systematics -- such as photometric blends, biased extinction values, or metallicity misestimation -- inflate the true uncertainties beyond the reported statistical errors. It also highlights the importance of accurate abundance measurements for subgiant ages (see also Appendix \ref{app:feh}). 
For the full \citet{Nataf2024} sample, the median absolute deviation scale is substantially smaller than the standard deviation ($\sigma_{z,\mathrm{MAD}}=1.22$ versus $\sigma_z=2.73$), indicating that the distribution is not uniformly broadened. Instead, it consists of a moderately broadened core plus a tail of larger discrepancies: 47 of the 61 systems ($77\%$) have $|z|<2$, while 14 systems ($23\%$) have $|z|>2$. By contrast, $\sigma_{z,\mathrm{MAD}} \approx \sigma_{z}$ for the \citet{XiangRix2022} and \citet{Wang2023} samples, indicating that their interpretations are not being driven by an outlier tail in the same way.

One potential contributor to this excess scatter is photometric variability, which can bias UV- and optical-based metallicity estimates if present (e.g., spot-induced variability with wavelength-dependent amplitudes). The \citet{Nataf2024} analysis mitigates this by excluding stars above the 95th percentile in Gaia G-band photometric variability. Among the wide-binary systems with large age discrepancies ($\sigma_z \gtrsim 2.5$), we find that only one source (Gaia DR3 2153929092538747904) exceeds this threshold of 0.95 (${\tt VarExcess}=0.99$) and exhibits clear periodic variability in ASAS-SN light curves. In contrast the other $3$ stars with $\sigma_z > 2.5$ and $0.90 < \mathrm{VarExcess} < 0.95$ show no detectable variability in ASAS-SN or ZTF data, supporting the effectiveness of the adopted variability cut. While photometric variability can potentially explain individual outliers, and should be considered for age estimation, the elevated dispersion in the Primary Sample persists even after excluding such sources. This indicates that additional systematics dominate the remaining discrepancy.

In summary, $\sigma_z \lsim 1$ for the \citet{XiangRix2022} and \citet{Wang2023} samples indicates that their reported age uncertainties are accurate, while $\sigma_z > 1$ for the \citet{Nataf2024} sample indicates that their errors are underestimated by a factor of $2-3$.
The reduction in scatter from the full \citet{Nataf2024} sample to their Primary Sample, and the improved agreement for high-SNR ($>50$) sources in \citet{Wang2023}, show that stringent quality cuts enhance reliability. 
Among the three age catalogs considered, \citet{XiangRix2022} achieve the most precise age estimates that also have empirically validated errors, as determined by wide binaries. The median fractional age uncertainty in their sample is $7.5\%$.
While the \citet{Wang2023} sample also has consistent errors for red giant and red clump stars, their median uncertainties are $\sim30\%$ ($\sim25\%$) for the total (SNR$>50$) sample. 
The success of \citet{XiangRix2022} reflects both the greater intrinsic age sensitivity of subgiants and the use of spectroscopic chemical abundance measurements to anchor their isochrone modeling.



\section{Discussion}\label{sec:discussion}

\subsection{Interpreting wide binary Constraints}\label{subsec:interp_wb_constraints}

The uncertainties inferred from our wide binary analysis represent a lower limit on the true total age uncertainties. Since both components share nearly identical metallicities and similar effective temperatures (when both are evolved), systematics that depend on these parameters are largely canceled out. Likewise, any global offset in the age scale (e.g., all ages being systematically over- or underestimated) would not manifest in our relative test.

These limitations imply that while wide binaries provide a useful test of reported uncertainties, they do not fully capture absolute accuracy. Systematic uncertainties associated with stellar evolution models set another floor for absolute age uncertainties \citep[e.g.][]{Morales25,Ying25}, which are crucial for cosmological and chemo-dynamical applications where absolute ages matter, such as in calibrating the age of the Universe or the early enrichment timeline. Our results therefore complement, rather than replace, absolute calibration efforts.

\subsection{Ages of Red Giants from [C/N] ratios}\label{subsec:red_giant_CN}
Another method of determining stellar ages for red giants relies on chemical clocks: elemental abundance ratios that trace stellar evolution or Galactic chemical enrichment. For red giants, the most widely used method for aging them is the surface carbon-to-nitrogen ratio ([C/N]), which changes during the first dredge-up and subsequent mixing on the RGB. This ratio correlates with stellar mass and therefore with age \citep[e.g.,][]{Masseron15,Martig16,Roberts24}. In contrast to abundance ratios such as [Fe/H] or [$\alpha$/Fe], which depend on Galactic chemical-evolution modeling and can introduce population-dependent biases \citep[e.g.,][]{Pagel09,Nissen15}, [C/N] is directly linked to stellar structure and internal mixing physics.

However, theoretical modeling of mixing and extra-mixing processes remains uncertain, particularly at low metallicities, where these effects can alter the [C/N]-mass relation \citep[e.g.,][]{Shetrone19,Roberts24}. Empirical calibration to asteroseismic masses and ages, for example, has proven valuable for anchoring this relation \citep[e.g.,][]{Bellinger16,Ness16, Mackereth19,Hon20,Anders23,Leung23,StoneMartinez24,StoneMartinez25}. Currently, these estimates provide reported age precisions of $\sim20-60\%$ \citep[e.g.,][]{Anders23,StoneMartinez25}. 

When cross-matched with our wide binary catalog, we find no binaries in which both components are present in the \citet{Anders23} APOGEE sample, while seven such pairs exist in the \citet{StoneMartinez25} SDSS-V DR19 catalog. For these systems, the inferred age differences are in fact consistent with the reported uncertainties ($\sigma_z = 1.5$), considering the small sample size, but the reported uncertainties are quite large, with a median value of $\approx28\%$. Thus, while the reported errors appear realistic, their true precision is comparable to that of isochrone ages for red giants \citep[e.g.,][]{Wang2023}.
The continued expansion of spectroscopic surveys, combined with large samples of asteroseismic masses and wide binaries (for calibrating uncertainties), offers an opportunity to refine these abundance-based age relations across metallicity and evolutionary phase.

\section{Conclusions}\label{sec:conclusions}
Precise stellar ages are crucial for reconstructing the Milky Way’s star formation, assembly, and chemical-enrichment history, as well as for testing stellar evolution models. As such, several recent studies estimated ages for millions of evolved stars throughout the Galaxy, reporting $\sim10\%$ uncertainties.
In this study, we use wide binaries as an empirical benchmark to test the reported age uncertainties of evolved stars. Because the two members of a wide binary are coeval, the average difference between their reported ages directly measures the true uncertainties on such estimates. 

Our analysis shows that the subgiant ages of \citet{XiangRix2022} and the red giant/red clump ages from \citet{Wang2023} are consistent with their quoted precisions ($\sigma_z\approx1$). The median fractional age uncertainty in the \citet{XiangRix2022} and \citet{Wang2023} sample is $7.5\%$ and $25-30\%$, respectively. 
In contrast, subgiant ages from the \citet{Nataf2024} catalog exhibit underestimated age uncertainties, with wide binary tests indicating effective errors $\sim2-3\times$ larger than reported formal errors. Even among their quality-controlled subsample, we find $\sigma_z\approx2$, demonstrating that additional systematics remain unaccounted for (Figure \ref{fig:RG_ages_accuracy}).

A key difference between the subgiant samples of \citet{XiangRix2022} and \citet{Nataf2024} lies in their treatment of chemical abundances. The former use spectroscopic measurements of both [Fe/H] and [$\alpha$/Fe], whereas the latter rely primarily on UV-sensitive photometric metallicities. This highlights the importance of accurate chemical abundances for reliable isochrone ages. For Galactic archaeology applications that demand high-precision ages, the \citet{XiangRix2022} subgiant catalog currently offers the most precise and empirically validated ages among those considered here. In general, we interpret the wide-binary comparison as a consistency test of the full age-estimation pipeline, rather than any single ingredient used in the calculation such as metallicities or choice of isochrone models (see also Appendix \ref{app:feh} and Figure \ref{fig:deltas_sigma_only}).

Beyond testing reported age precisions, this study also underscores the utility of wide binaries as a model-independent benchmark for stellar age measurements. Unlike internal validation tests or calibrations to clusters, wide binaries provide a large set of validation experiments across the Galaxy, and most importantly, across a wide range of ages, metallicities, and $\alpha$-abundances, enabling empirical assessments of true uncertainties.

{\it Gaia} XP spectra provide metallicities for hundreds of millions of stars \citep[e.g.,][]{Andrae23}, which, when combined with effective temperatures and luminosities, may allow ages to be estimated for subgiants without the need for higher-resolution spectroscopy. Extending such XP-based analyses and validating them with wide binaries represents a promising avenue for providing isochrone ages for a larger sample of subgiants. Looking ahead, as surveys deliver larger samples of evolved stars with precise chemical abundances, wide binaries will remain a stringent benchmark for calibrating the precision of stellar age and abundance measurements.

\section{Acknowledgments} \label{acknowledgments}
We thank two anonymous referees for providing constructive feedback that improved the manuscript. We thank Chun Wang and Yang Huang for useful discussions.

C.S. acknowledges support from the Department of Energy Computational Science Graduate Fellowship.
This material is based upon work supported by the U.S. Department of Energy, Office of Science, Office of Advanced Scientific Computing Research, under Award Number DE-SC0026073. This research was supported by NSF grant AST-2307232 and by Scialog grant \#SA-LSST-2024-084a from the
Research Corporation for Science Advancement and the Heising-Simons Foundation.

This work presents results from the European Space Agency (ESA) space mission {\it Gaia}. {\it Gaia} data are being processed by the {\it Gaia} Data Processing and Analysis Consortium (DPAC). Funding for the DPAC is provided by national institutions, in particular the institutions participating in the {\it Gaia} MultiLateral Agreement (MLA). The {\it Gaia} mission website is \url{https://www.cosmos.esa.int/gaia}. The {\it Gaia} archive website is \url{https://archives.esac.esa.int/gaia}.

\textit{Software.} This work made use of the following software packages: \texttt{Jupyter} \citep{2007CSE.....9c..21P,kluyver2016jupyter}, \texttt{matplotlib} \citep{Hunter:2007}, \texttt{pandas} \citep{mckinney-proc-scipy-2010}, \texttt{python} \citep{python}, and \texttt{TOPCAT} \citep{2005ASPC..347...29T}.
This work made use of \texttt{OverCite} \citep{Shariat2026}, an in-editor citation tool for \LaTeX.
Software citation information aggregated using \texttt{\href{https://www.tomwagg.com/software-citation-station/}{The Software Citation Station}} \citep{software-citation-station-paper,software-citation-station-zenodo}.

\appendix 
\section{Age and Metallicity}\label{app:feh}
\begin{figure*}[h]
\centering
\includegraphics[width=0.99\textwidth]{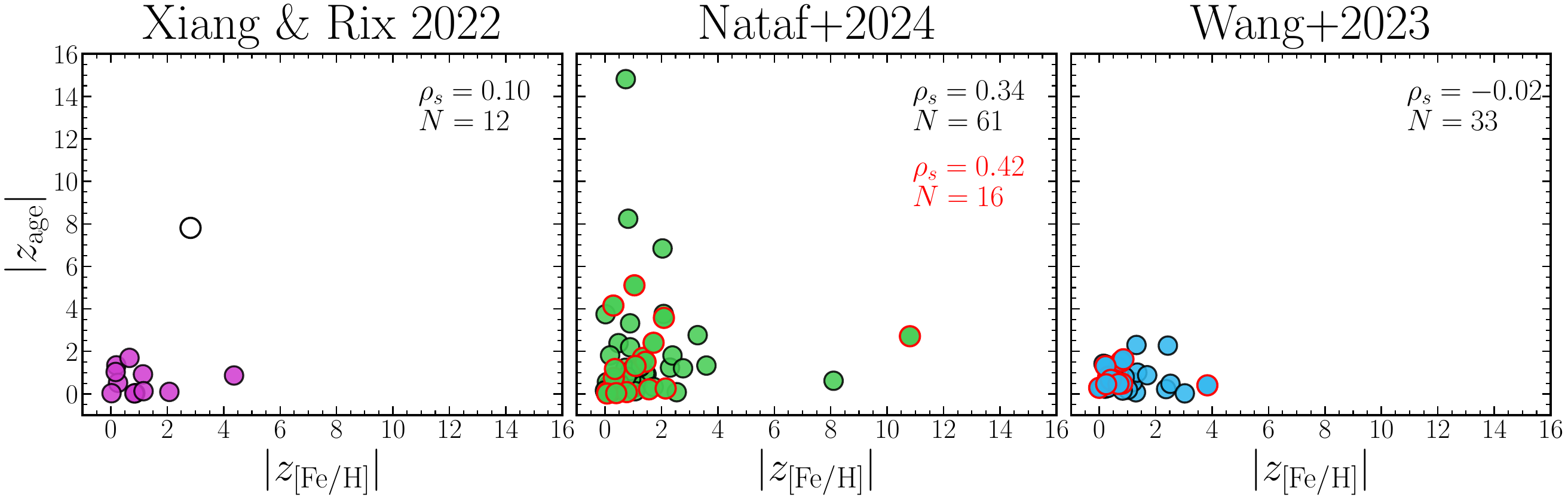}
\caption{
Normalized metallicity and age differences for wide-binary pairs in the \citet{XiangRix2022}, \citet{Nataf2024}, and \citet{Wang2023} samples. Here $|z_{[{\rm Fe/H}]}|$ and $|z_{\rm age}|$ denote the absolute value of the component-to-component differences divided by their quoted uncertainties. Values near zero indicate consistency within the formal error. The formatting is the same as Figure \ref{fig:RG_ages_accuracy}. The annotated $\rho_s$ values are the Spearman rank correlation between $|z_{[{\rm Fe/H}]}|$ and $|z_{\rm age}|$, with $N$ the number of systems in each panel. Age and metallicity uncertainties for the \citet{XiangRix2022} and \citet{Wang2023} sample are generally uncorrelated, while the \citet{Nataf2024} sample shows moderate correlation between them.
}
\label{fig:deltas_sigma_only}
\end{figure*}

We test whether systems that appear discrepant in age also tend to be discrepant in metallicity. Figure~\ref{fig:deltas_sigma_only} compares the normalized component-to-component metallicity differences, $|z_{[{\rm Fe/H}]}|$, with the corresponding normalized age differences, $|z_{\rm age}|$ (Equation \ref{eq:z}), for the three catalogs considered in this work. If the quoted uncertainties fully capture the observational and modeling error budget, the points should cluster near the origin (0,0) with a characteristic spread of order unity. Any correlation between $|z_{[{\rm Fe/H}]}|$ and $|z_{\rm age}|$ would suggest that age inconsistencies are at least partly linked to systems with metallicity inconsistencies.

The three samples show qualitatively different behavior. The \citet{Wang2023} sample remains tightly clustered near the origin and shows essentially no correlation between normalized metallicity and age discrepancies ($\rho_s=-0.02$). This is consistent with our broader conclusion that the Wang uncertainties are comparatively realistic, even if the ages themselves are substantially less precise than those of \citet{XiangRix2022}. The \citet{XiangRix2022} sample likewise does not exhibit a strong global trend ($\rho_s=0.10$), although the small sample size means that individual systems can still have substantial leverage. In particular, the most conspicuous \citet{XiangRix2022} outlier remains strongly inconsistent in age while also showing a noticeable metallicity discrepancy, but the sample as a whole does not support a clear monotonic coupling between metallicity mismatch and age mismatch.

The \citet{Nataf2024} sample shows the strongest trend, with a modest positive correlation ($\rho_s=0.34$ for all, $\rho_s=0.42$ for the primary sample). Systems that are more discrepant in metallicity also tend, on average, to be more discrepant in age. This does not imply that metallicity errors alone explain the full age scatter. Instead, it likely suggests that part of the broadening seen in their age comparison reflects a more general underestimation of the total uncertainty budget in at least some systems. At the same time, the correlation is not tight, and many systems with small metallicity discrepancies still exhibit non-negligible age offsets. We therefore interpret Figure~\ref{fig:deltas_sigma_only} as supporting the same overall picture found throughout this paper: Wang appears comparatively well calibrated, \citet{XiangRix2022} is globally precise but sensitive to a small number of problematic systems, and \citet{Nataf2024} shows a broader distribution in which the formal uncertainties are too small for at least a non-negligible subset of the sample.

\clearpage
\onecolumngrid

\section{Wide Binary Table}\label{app:wbs}
Table~\ref{tab:wbs} shows a subset of the wide-binary catalog used in this work. The full machine-readable version is available at \url{https://zenodo.org/records/17957444}.

\begingroup
\scriptsize
\setlength{\tabcolsep}{2.5pt}
\renewcommand{\arraystretch}{1.08}
\setlength{\LTcapwidth}{0.98\textwidth}
\renewenvironment{deluxetable}[1]{\begin{longtable}{@{}#1@{}}}{\end{longtable}}
\renewcommand{\tablecaption}[1]{\caption{#1}\\[0.5ex]\toprule}
\renewcommand{\tablehead}[1]{#1\\\midrule\endfirsthead}
\renewcommand{\startdata}{}
\renewcommand{\enddata}{}
\begin{deluxetable}{lcccccccc}
    \tablecaption{{\it Gaia} parameters of Wide  Binaries\label{tab:wbs}. Subscript `1' refers to the primary (brighter) and `2' refers to the secondary (fainter) companion. Entries marked with `*' denote additional quality cuts (see text).}
    \tablehead{
    \colhead{Catalog} &
    \colhead{{\it Gaia} DR3 ID 1} &
    \colhead{{\it Gaia} DR3 ID 2} &  
    \colhead{$\varpi_1$} & 
    \colhead{$\varpi_2$} & 
    \colhead{$s$} &
    \colhead{Age 1} &
    \colhead{Age 2} &
    \colhead{$\mathcal{R}$} 
    \\
    \colhead{} &
    \colhead{} &
    \colhead{} &  
    \colhead{[mas]} & 
    \colhead{[mas]} &
    \colhead{[au]} &
    \colhead{[Gyr]} &
    \colhead{[Gyr]} &
    \colhead{}
    }
    \startdata
        XR22\tablenotemark{a} & 2684790761174460032 & 2684790658095155840 & $1.238^{+0.017}_{-0.017}$ & $1.247^{+0.015}_{-0.015}$ & 17590 & $3.85^{+0.14}_{-0.14}$ & $5.15^{+0.09}_{-0.09}$ & 0.000284 \\
XR22 & 949512025266986368 & 949511995203757568 & $3.210^{+0.014}_{-0.014}$ & $3.213^{+0.013}_{-0.013}$ & 10407 & $5.26^{+0.25}_{-0.25}$ & $5.31^{+0.46}_{-0.46}$ & 0.000495 \\
XR22 & 1331136194689123072 & 1331136194689123200 & $2.046^{+0.010}_{-0.010}$ & $2.048^{+0.010}_{-0.010}$ & 10348 & $11.98^{+0.65}_{-0.65}$ & $11.31^{+0.42}_{-0.42}$ & 0.000000 \\
XR22 & 656616764883595264 & 656616769173736960 & $1.192^{+0.015}_{-0.015}$ & $1.180^{+0.015}_{-0.015}$ & 7201 & $5.20^{+0.20}_{-0.20}$ & $4.82^{+0.20}_{-0.20}$ & 0.000056 \\
XR22 & 774012622101896832 & 774012622101896960 & $0.971^{+0.019}_{-0.019}$ & $0.945^{+0.019}_{-0.019}$ & 10004 & $11.61^{+0.55}_{-0.55}$ & $10.92^{+0.51}_{-0.51}$ & 0.000000 \\
XR22 & 2098638466710584064 & 2098638466710583552 & $0.958^{+0.012}_{-0.012}$ & $0.934^{+0.012}_{-0.012}$ & 24694 & $8.91^{+0.28}_{-0.28}$ & $8.11^{+0.38}_{-0.38}$ & 0.001709 \\
XR22 & 742452824453216000 & 742452824453215872 & $0.924^{+0.020}_{-0.020}$ & $0.913^{+0.021}_{-0.021}$ & 6987 & $10.83^{+0.41}_{-0.41}$ & $10.80^{+0.78}_{-0.78}$ & 0.000000 \\
XR22 & 3283076097735358592 & 3283076097735358464 & $0.916^{+0.015}_{-0.015}$ & $0.877^{+0.016}_{-0.016}$ & 13466 & $4.55^{+0.16}_{-0.16}$ & $4.70^{+0.24}_{-0.24}$ & 0.000269 \\
XR22 & 605026206128002432 & 605026304912470272 & $0.871^{+0.017}_{-0.017}$ & $0.879^{+0.020}_{-0.020}$ & 122214 & $7.55^{+0.52}_{-0.52}$ & $7.53^{+0.76}_{-0.76}$ & 0.000272 \\
XR22 & 1271083343481863296 & 1271083343481863552 & $0.747^{+0.018}_{-0.018}$ & $0.753^{+0.017}_{-0.017}$ & 14891 & $10.32^{+0.49}_{-0.49}$ & $10.30^{+0.38}_{-0.38}$ & 0.000000 \\
XR22 & 139325960240584448 & 139325960240584320 & $0.650^{+0.023}_{-0.023}$ & $0.691^{+0.024}_{-0.024}$ & 16145 & $3.35^{+0.36}_{-0.36}$ & $3.78^{+0.21}_{-0.21}$ & 0.000180 \\
XR22 & 3712569838037565568 & 3712569838037565696 & $0.624^{+0.026}_{-0.026}$ & $0.618^{+0.028}_{-0.028}$ & 17012 & $13.17^{+1.16}_{-1.16}$ & $12.96^{+1.15}_{-1.15}$ & 0.000085 \\
W23* & 2806661695148871808 & 2806661695148871680 & $2.119^{+0.015}_{-0.015}$ & $2.176^{+0.015}_{-0.015}$ & 4896 & $3.65^{+1.00}_{-1.00}$ & $1.89^{+0.57}_{-0.57}$ & 0.000033 \\
W23* & 611480030145802496 & 611480030144420608 & $1.826^{+0.022}_{-0.022}$ & $1.777^{+0.017}_{-0.017}$ & 1143 & $5.07^{+1.30}_{-1.30}$ & $3.76^{+0.92}_{-0.92}$ & 0.000001 \\
W23* & 3844873556811289600 & 3844873556811289472 & $1.825^{+0.014}_{-0.014}$ & $1.825^{+0.015}_{-0.015}$ & 4762 & $5.04^{+1.24}_{-1.24}$ & $4.11^{+0.86}_{-0.86}$ & 0.000001 \\
W23* & 590400742973158528 & 590400742973157760 & $1.806^{+0.016}_{-0.016}$ & $1.749^{+0.018}_{-0.018}$ & 7474 & $3.32^{+0.69}_{-0.69}$ & $4.03^{+0.93}_{-0.93}$ & 0.000259 \\
\midrule
\multicolumn{9}{@{}l@{}}{\parbox{0.96\textwidth}{\footnotesize\itshape The full table is published in machine-readable form at \url{https://zenodo.org/records/17957444}.}} \\
\bottomrule
\enddata
    \enddata
    
\end{deluxetable}
\par\footnotesize $^a$ Not used in statistical analysis.
\par\footnotesize Columns: (1) Catalog name among \citealt{XiangRix2022} (XR22), \citealt{Wang2023} (W23), \citealt{Nataf2024} (N24); ($2-3$) {\it Gaia} DR3 source ID of the primary and secondary; ($4-5$) Parallaxes of the primary and secondary in mas; (6) Projected separation of the binary in au; ($7-8$) Isochrone ages for the primary and secondary, (9) Chance alignment probability of the binary \citep[$\mathcal{R}$; see][]{EB21_widebin}.
\endgroup

\twocolumngrid

\bibliography{references}

@ARTICLE{2007CSE.....9c..21P,
       author = {{Perez}, Fernando and {Granger}, Brian E.},
        title = "{IPython: A System for Interactive Scientific Computing}",
      journal = {Computing in Science and Engineering},
         year = "2007",
        month = "Jan",
       volume = {9},
       number = {3},
        pages = {21-29},
          doi = {10.1109/MCSE.2007.53},
       adsurl = {https://ui.adsabs.harvard.edu/abs/2007CSE.....9c..21P}
}

@conference{kluyver2016jupyter,
    title = {Jupyter Notebooks -- a publishing format for reproducible computational workflows},
    author = {Thomas Kluyver and Benjamin Ragan-Kelley and Fernando P{\'e}rez and Brian Granger and Matthias Bussonnier and Jonathan Frederic and Kyle Kelley and Jessica Hamrick and Jason Grout and Sylvain Corlay and Paul Ivanov and Dami{\'a}n Avila and Safia Abdalla and Carol Willing},
    booktitle = {Positioning and Power in Academic Publishing: Players, Agents and Agendas},
    editor = {F. Loizides and B. Schmidt},
    organization = {IOS Press},
    pages = {87 - 90},
    year = {2016}
}

@Article{Hunter:2007,
  Author    = {Hunter, J. D.},
  Title     = {Matplotlib: A 2D graphics environment},
  Journal   = {Computing in Science \& Engineering},
  Volume    = {9},
  Number    = {3},
  Pages     = {90--95},
  publisher = {IEEE COMPUTER SOC},
  doi       = {10.1109/MCSE.2007.55},
  year      = 2007
}

@InProceedings{mckinney-proc-scipy-2010,
  author    = { {W}es {M}c{K}inney },
  title     = { {D}ata {S}tructures for {S}tatistical {C}omputing in {P}ython },
  booktitle = { {P}roceedings of the 9th {P}ython in {S}cience {C}onference },
  pages     = { 56 - 61 },
  year      = { 2010 },
  editor    = { {S}t\'efan van der {W}alt and {J}arrod {M}illman },
  doi       = { 10.25080/Majora-92bf1922-00a }
}

@book{python,
  author    = {Van Rossum, Guido and Drake, Fred L.},
  title     = {Python 3 Reference Manual},
  year      = {2009},
  isbn      = {1441412697},
  publisher = {CreateSpace},
  address   = {Scotts Valley, CA}
}

@INPROCEEDINGS{2005ASPC..347...29T,
       author = {{Taylor}, M.~B.},
        title = "{TOPCAT \& STIL: Starlink Table/VOTable Processing Software}",
    booktitle = {Astronomical Data Analysis Software and Systems XIV},
         year = 2005,
       editor = {{Shopbell}, P. and {Britton}, M. and {Ebert}, R.},
       series = {Astronomical Society of the Pacific Conference Series},
       volume = {347},
        month = dec,
        pages = {29},
       adsurl = {https://ui.adsabs.harvard.edu/abs/2005ASPC..347...29T}
}

@ARTICLE{software-citation-station-paper,
       author = {{Wagg}, Tom and {Broekgaarden}, Floor S.},
        title = "{Streamlining and standardizing software citations with The Software Citation Station}",
      journal = {arXiv e-prints},
         year = 2024,
        month = jun,
          eid = {arXiv:2406.04405},
        pages = {arXiv:2406.04405},
archivePrefix = {arXiv},
       eprint = {2406.04405},
 primaryClass = {astro-ph.IM},
       adsurl = {https://ui.adsabs.harvard.edu/abs/2024arXiv240604405W}
}

@software{software-citation-station-zenodo,
  author       = {Tom Wagg and
                  Floor Broekgaarden and
                  Phil Van-Lane and
                  Kai Wu and
                  Kayhan Gültekin},
  title        = {TomWagg/software-citation-station: v1.4},
  month        = nov,
  year         = 2025,
  publisher    = {Zenodo},
  version      = {v1.4},
  doi          = {10.5281/zenodo.17654855},
  url          = {https://doi.org/10.5281/zenodo.17654855},
  swhid        = {swh:1:dir:9a009430037c791424a572f542e9a5d5c1fb44ff
                   ;origin=https://doi.org/10.5281/zenodo.13225526;vi
                   sit=swh:1:snp:ef11f058d718d691f0661c9445c2251328cb
                   ac95;anchor=swh:1:rel:84cde4e532032537b7f014c4467c
                   102430a53fa2;path=TomWagg-software-citation-
                   station-61a588a
                  },
}

@ARTICLE{EB21_widebin,
       author = {{El-Badry}, Kareem and {Rix}, Hans-Walter and {Heintz}, Tyler M.},
        title = "{A million binaries from Gaia eDR3: sample selection and validation of Gaia parallax uncertainties}",
      journal = {\mnras},
         year = 2021,
        month = sep,
       volume = {506},
       number = {2},
        pages = {2269-2295},
          doi = {10.1093/mnras/stab323},
archivePrefix = {arXiv},
       eprint = {2101.05282},
 primaryClass = {astro-ph.SR},
       adsurl = {https://ui.adsabs.harvard.edu/abs/2021MNRAS.506.2269E}
}

@ARTICLE{EB24_review,
       author = {{El-Badry}, Kareem},
        title = "{Gaia's binary star renaissance}",
      journal = {\nar},
         year = 2024,
        month = jun,
       volume = {98},
          eid = {101694},
        pages = {101694},
          doi = {10.1016/j.newar.2024.101694},
archivePrefix = {arXiv},
       eprint = {2403.12146},
 primaryClass = {astro-ph.SR},
       adsurl = {https://ui.adsabs.harvard.edu/abs/2024NewAR..9801694E}
}

@ARTICLE{Lepine07,
       author = {{L{\'e}pine}, S{\'e}bastien and {Bongiorno}, Bethany},
        title = "{New Distant Companions to Known Nearby Stars. II. Faint Companions of Hipparcos Stars and the Frequency of Wide Binary Systems}",
      journal = {\aj},
         year = 2007,
        month = mar,
       volume = {133},
       number = {3},
        pages = {889-905},
          doi = {10.1086/510333},
archivePrefix = {arXiv},
       eprint = {astro-ph/0610605},
 primaryClass = {astro-ph},
       adsurl = {https://ui.adsabs.harvard.edu/abs/2007AJ....133..889L}
}

@ARTICLE{Bressan12,
       author = {{Bressan}, Alessandro and {Marigo}, Paola and {Girardi}, L{\'e}o. and {Salasnich}, Bernardo and {Dal Cero}, Claudia and {Rubele}, Stefano and {Nanni}, Ambra},
        title = "{PARSEC: stellar tracks and isochrones with the PAdova and TRieste Stellar Evolution Code}",
      journal = {\mnras},
         year = 2012,
        month = nov,
       volume = {427},
       number = {1},
        pages = {127-145},
          doi = {10.1111/j.1365-2966.2012.21948.x},
archivePrefix = {arXiv},
       eprint = {1208.4498},
 primaryClass = {astro-ph.SR},
       adsurl = {https://ui.adsabs.harvard.edu/abs/2012MNRAS.427..127B}
}

@ARTICLE{XiangRix2022,
       author = {{Xiang}, Maosheng and {Rix}, Hans-Walter},
        title = "{A time-resolved picture of our Milky Way's early formation history}",
      journal = {\nat},
         year = 2022,
        month = mar,
       volume = {603},
       number = {7902},
        pages = {599-603},
          doi = {10.1038/s41586-022-04496-5},
archivePrefix = {arXiv},
       eprint = {2203.12110},
 primaryClass = {astro-ph.GA},
       adsurl = {https://ui.adsabs.harvard.edu/abs/2022Natur.603..599X}
}

@ARTICLE{Nataf2024,
       author = {{Nataf}, David M. and {Schlaufman}, Kevin C. and {Reggiani}, Henrique and {Hahn}, Isabel},
        title = "{Accurate, Precise, and Physically Self-consistent Ages and Metallicities for 400,000 Solar Neighborhood Subgiant Branch Stars}",
      journal = {\apj},
         year = 2024,
        month = nov,
       volume = {976},
       number = {1},
          eid = {87},
        pages = {87},
          doi = {10.3847/1538-4357/ad7c4e},
archivePrefix = {arXiv},
       eprint = {2407.18307},
 primaryClass = {astro-ph.SR},
       adsurl = {https://ui.adsabs.harvard.edu/abs/2024ApJ...976...87N}
}

@ARTICLE{Wang2023,
       author = {{Wang}, Chun and {Huang}, Yang and {Zhou}, Yutao and {Zhang}, Huawei},
        title = "{Precise masses and ages of  1 million RGB and RC stars observed by LAMOST}",
      journal = {\aap},
         year = 2023,
        month = jul,
       volume = {675},
          eid = {A26},
        pages = {A26},
          doi = {10.1051/0004-6361/202245809},
archivePrefix = {arXiv},
       eprint = {2305.04528},
 primaryClass = {astro-ph.GA},
       adsurl = {https://ui.adsabs.harvard.edu/abs/2023A&A...675A..26W}
}

@ARTICLE{Soderblom2010,
       author = {{Soderblom}, David R.},
        title = "{The Ages of Stars}",
      journal = {\araa},
         year = 2010,
        month = sep,
       volume = {48},
        pages = {581-629},
          doi = {10.1146/annurev-astro-081309-130806},
archivePrefix = {arXiv},
       eprint = {1003.6074},
 primaryClass = {astro-ph.SR},
       adsurl = {https://ui.adsabs.harvard.edu/abs/2010ARA&A..48..581S}
}

@ARTICLE{Twarog80,
       author = {{Twarog}, B.~A.},
        title = "{The chemical evolution of the solar neighborhood. II - The age-metallicity relation and the history of star formation in the galactic disk}",
      journal = {\apj},
         year = 1980,
        month = nov,
       volume = {242},
        pages = {242-259},
          doi = {10.1086/158460},
       adsurl = {https://ui.adsabs.harvard.edu/abs/1980ApJ...242..242T}
}

@ARTICLE{Casagrande16,
       author = {{Casagrande}, L. and {Silva Aguirre}, V. and {Schlesinger}, K.~J. and {Stello}, D. and {Huber}, D. and {Serenelli}, A.~M. and {Sch{\"o}nrich}, R. and {Cassisi}, S. and {Pietrinferni}, A. and {Hodgkin}, S. and {Milone}, A.~P. and {Feltzing}, S. and {Asplund}, M.},
        title = "{Measuring the vertical age structure of the Galactic disc using asteroseismology and SAGA}",
      journal = {\mnras},
         year = 2016,
        month = jan,
       volume = {455},
       number = {1},
        pages = {987-1007},
          doi = {10.1093/mnras/stv2320},
archivePrefix = {arXiv},
       eprint = {1510.01376},
 primaryClass = {astro-ph.GA},
       adsurl = {https://ui.adsabs.harvard.edu/abs/2016MNRAS.455..987C}
}

@ARTICLE{Yi01,
       author = {{Yi}, Sukyoung and {Demarque}, Pierre and {Kim}, Yong-Cheol and {Lee}, Young-Wook and {Ree}, Chang H. and {Lejeune}, Thibault and {Barnes}, Sydney},
        title = "{Toward Better Age Estimates for Stellar Populations: The Y$^{2}$ Isochrones for Solar Mixture}",
      journal = {\apjs},
         year = 2001,
        month = oct,
       volume = {136},
       number = {2},
        pages = {417-437},
          doi = {10.1086/321795},
archivePrefix = {arXiv},
       eprint = {astro-ph/0104292},
 primaryClass = {astro-ph},
       adsurl = {https://ui.adsabs.harvard.edu/abs/2001ApJS..136..417Y}
}

@ARTICLE{Demarque04,
       author = {{Demarque}, Pierre and {Woo}, Jong-Hak and {Kim}, Yong-Cheol and {Yi}, Sukyoung K.},
        title = "{Y$^{2}$ Isochrones with an Improved Core Overshoot Treatment}",
      journal = {\apjs},
         year = 2004,
        month = dec,
       volume = {155},
       number = {2},
        pages = {667-674},
          doi = {10.1086/424966},
       adsurl = {https://ui.adsabs.harvard.edu/abs/2004ApJS..155..667D}
}

@ARTICLE{Bedding11,
       author = {{Bedding}, Timothy R. and {Mosser}, Benoit and {Huber}, Daniel and {Montalb{\'a}n}, Josefina and {Beck}, Paul and {Christensen-Dalsgaard}, J{\o}rgen and {Elsworth}, Yvonne P. and {Garc{\'\i}a}, Rafael A. and {Miglio}, Andrea and {Stello}, Dennis and {White}, Timothy R. and {De Ridder}, Joris and {Hekker}, Saskia and {Aerts}, Conny and {Barban}, Caroline and {Belkacem}, Kevin and {Broomhall}, Anne-Marie and {Brown}, Timothy M. and {Buzasi}, Derek L. and {Carrier}, Fabien and {Chaplin}, William J. and {di Mauro}, Maria Pia and {Dupret}, Marc-Antoine and {Frandsen}, S{\o}ren and {Gilliland}, Ronald L. and {Goupil}, Marie-Jo and {Jenkins}, Jon M. and {Kallinger}, Thomas and {Kawaler}, Steven and {Kjeldsen}, Hans and {Mathur}, Savita and {Noels}, Arlette and {Silva Aguirre}, Victor and {Ventura}, Paolo},
        title = "{Gravity modes as a way to distinguish between hydrogen- and helium-burning red giant stars}",
      journal = {\nat},
         year = 2011,
        month = mar,
       volume = {471},
       number = {7340},
        pages = {608-611},
          doi = {10.1038/nature09935},
archivePrefix = {arXiv},
       eprint = {1103.5805},
 primaryClass = {astro-ph.SR},
       adsurl = {https://ui.adsabs.harvard.edu/abs/2011Natur.471..608B}
}

@ARTICLE{Stello13,
       author = {{Stello}, Dennis and {Huber}, Daniel and {Bedding}, Timothy R. and {Benomar}, Othman and {Bildsten}, Lars and {Elsworth}, Yvonne P. and {Gilliland}, Ronald L. and {Mosser}, Beno{\^\i}t and {Paxton}, Bill and {White}, Timothy R.},
        title = "{Asteroseismic Classification of Stellar Populations among 13,000 Red Giants Observed by Kepler}",
      journal = {\apjl},
         year = 2013,
        month = mar,
       volume = {765},
       number = {2},
          eid = {L41},
        pages = {L41},
          doi = {10.1088/2041-8205/765/2/L41},
archivePrefix = {arXiv},
       eprint = {1302.0858},
 primaryClass = {astro-ph.SR},
       adsurl = {https://ui.adsabs.harvard.edu/abs/2013ApJ...765L..41S}
}

@ARTICLE{Pinsonneault14,
       author = {{Pinsonneault}, Marc H. and {Elsworth}, Yvonne and {Epstein}, Courtney and {Hekker}, Saskia and {M{\'e}sz{\'a}ros}, Sz. and {Chaplin}, William J. and {Johnson}, Jennifer A. and {Garc{\'\i}a}, Rafael A. and {Holtzman}, Jon and {Mathur}, Savita and {Garc{\'\i}a P{\'e}rez}, Ana and {Silva Aguirre}, Victor and {Girardi}, L{\'e}o and {Basu}, Sarbani and {Shetrone}, Matthew and {Stello}, Dennis and {Allende Prieto}, Carlos and {An}, Deokkeun and {Beck}, Paul and {Beers}, Timothy C. and {Bizyaev}, Dmitry and {Bloemen}, Steven and {Bovy}, Jo and {Cunha}, Katia and {De Ridder}, Joris and {Frinchaboy}, Peter M. and {Garc{\'\i}a-Hern{\'a}ndez}, D.~A. and {Gilliland}, Ronald and {Harding}, Paul and {Hearty}, Fred R. and {Huber}, Daniel and {Ivans}, Inese and {Kallinger}, Thomas and {Majewski}, Steven R. and {Metcalfe}, Travis S. and {Miglio}, Andrea and {Mosser}, Benoit and {Muna}, Demitri and {Nidever}, David L. and {Schneider}, Donald P. and {Serenelli}, Aldo and {Smith}, Verne V. and {Tayar}, Jamie and {Zamora}, Olga and {Zasowski}, Gail},
        title = "{The APOKASC Catalog: An Asteroseismic and Spectroscopic Joint Survey of Targets in the Kepler Fields}",
      journal = {\apjs},
         year = 2014,
        month = dec,
       volume = {215},
       number = {2},
          eid = {19},
        pages = {19},
          doi = {10.1088/0067-0049/215/2/19},
archivePrefix = {arXiv},
       eprint = {1410.2503},
 primaryClass = {astro-ph.SR},
       adsurl = {https://ui.adsabs.harvard.edu/abs/2014ApJS..215...19P}
}

@ARTICLE{Vrard16,
       author = {{Vrard}, M. and {Mosser}, B. and {Samadi}, R.},
        title = "{Period spacings in red giants. II. Automated measurement}",
      journal = {\aap},
         year = 2016,
        month = apr,
       volume = {588},
          eid = {A87},
        pages = {A87},
          doi = {10.1051/0004-6361/201527259},
archivePrefix = {arXiv},
       eprint = {1602.04940},
 primaryClass = {astro-ph.SR},
       adsurl = {https://ui.adsabs.harvard.edu/abs/2016A&A...588A..87V}
}

@ARTICLE{Elsworth17,
       author = {{Elsworth}, Yvonne and {Hekker}, Saskia and {Basu}, Sarbani and {Davies}, Guy R.},
        title = "{A new method for the asteroseismic determination of the evolutionary state of red-giant stars}",
      journal = {\mnras},
         year = 2017,
        month = apr,
       volume = {466},
       number = {3},
        pages = {3344-3352},
          doi = {10.1093/mnras/stw3288},
archivePrefix = {arXiv},
       eprint = {1612.04751},
 primaryClass = {astro-ph.SR},
       adsurl = {https://ui.adsabs.harvard.edu/abs/2017MNRAS.466.3344E}
}

@ARTICLE{Wu19,
       author = {{Wu}, Yaqian and {Xiang}, Maosheng and {Zhao}, Gang and {Bi}, Shaolan and {Liu}, Xiaowei and {Shi}, Jianrong and {Huang}, Yang and {Yuan}, Haibo and {Wang}, Chun and {Chen}, Bingqiu and {Huo}, Zhiying and {Ren}, Juanjuan and {Tian}, Zhijia and {Liu}, Kang and {Zhang}, Xianfei and {Li}, Yaguang and {Zhang}, Jinghua},
        title = "{Ages and masses of 0.64 million red giant branch stars from the LAMOST Galactic Spectroscopic Survey}",
      journal = {\mnras},
         year = 2019,
        month = apr,
       volume = {484},
       number = {4},
        pages = {5315-5329},
          doi = {10.1093/mnras/stz256},
archivePrefix = {arXiv},
       eprint = {1901.07233},
 primaryClass = {astro-ph.SR},
       adsurl = {https://ui.adsabs.harvard.edu/abs/2019MNRAS.484.5315W}
}

@ARTICLE{Ting18,
       author = {{Ting}, Yuan-Sen and {Hawkins}, Keith and {Rix}, Hans-Walter},
        title = "{A Large and Pristine Sample of Standard Candles across the Milky Way: {\ensuremath{\sim}}100,000 Red Clump Stars with 3\% Contamination}",
      journal = {\apjl},
         year = 2018,
        month = may,
       volume = {858},
       number = {1},
          eid = {L7},
        pages = {L7},
          doi = {10.3847/2041-8213/aabf8e},
archivePrefix = {arXiv},
       eprint = {1803.06650},
 primaryClass = {astro-ph.SR},
       adsurl = {https://ui.adsabs.harvard.edu/abs/2018ApJ...858L...7T}
}

@software{Morton15,
       author = {{Morton}, Timothy D.},
        title = "{isochrones: Stellar model grid package}",
 howpublished = {Astrophysics Source Code Library, record ascl:1503.010},
         year = 2015,
        month = mar,
          eid = {ascl:1503.010},
archivePrefix = {ascl},
       eprint = {1503.010},
       adsurl = {https://ui.adsabs.harvard.edu/abs/2015ascl.soft03010M}
}

@ARTICLE{Bonaca20,
       author = {{Bonaca}, Ana and {Conroy}, Charlie and {Cargile}, Phillip A. and {Naidu}, Rohan P. and {Johnson}, Benjamin D. and {Zaritsky}, Dennis and {Ting}, Yuan-Sen and {Caldwell}, Nelson and {Han}, Jiwon Jesse and {van Dokkum}, Pieter},
        title = "{Timing the Early Assembly of the Milky Way with the H3 Survey}",
      journal = {\apjl},
         year = 2020,
        month = jul,
       volume = {897},
       number = {1},
          eid = {L18},
        pages = {L18},
          doi = {10.3847/2041-8213/ab9caa},
archivePrefix = {arXiv},
       eprint = {2004.11384},
 primaryClass = {astro-ph.GA},
       adsurl = {https://ui.adsabs.harvard.edu/abs/2020ApJ...897L..18B}
}

@ARTICLE{Xiang17,
       author = {{Xiang}, Maosheng and {Liu}, Xiaowei and {Shi}, Jianrong and {Yuan}, Haibo and {Huang}, Yang and {Chen}, Bingqiu and {Wang}, Chun and {Tian}, Zhijia and {Wu}, Yaqian and {Yang}, Yong and {Zhang}, Huawei and {Huo}, Zhiying and {Ren}, Juanjuan},
        title = "{The Ages and Masses of a Million Galactic-disk Main-sequence Turnoff and Subgiant Stars from the LAMOST Galactic Spectroscopic Surveys}",
      journal = {\apjs},
         year = 2017,
        month = sep,
       volume = {232},
       number = {1},
          eid = {2},
        pages = {2},
          doi = {10.3847/1538-4365/aa80e4},
archivePrefix = {arXiv},
       eprint = {1707.06236},
 primaryClass = {astro-ph.SR},
       adsurl = {https://ui.adsabs.harvard.edu/abs/2017ApJS..232....2X}
}

@ARTICLE{Wang25,
       author = {{Wang}, Jia-Hui and {Xiang}, Maosheng and {Zhang}, Meng and {Xie}, Ji-Wei and {Ge}, Jian and {Zhang}, Jinghua and {Mou}, Lanya and {Liu}, Ji-Feng},
        title = "{Spectroscopic Ages for 4 Million Main-sequence Dwarf Stars from LAMOST DR10 Estimated with a Data-driven Approach}",
      journal = {\apjs},
         year = 2025,
        month = sep,
       volume = {280},
       number = {1},
          eid = {13},
        pages = {13},
          doi = {10.3847/1538-4365/aded16},
archivePrefix = {arXiv},
       eprint = {2508.03019},
 primaryClass = {astro-ph.SR},
       adsurl = {https://ui.adsabs.harvard.edu/abs/2025ApJS..280...13W}
}

@ARTICLE{Gruner23,
       author = {{Gruner}, D. and {Barnes}, S.~A. and {Janes}, K.~A.},
        title = "{Wide binaries demonstrate the consistency of rotational evolution between open cluster and field stars}",
      journal = {\aap},
         year = 2023,
        month = jul,
       volume = {675},
          eid = {A180},
        pages = {A180},
          doi = {10.1051/0004-6361/202346590},
archivePrefix = {arXiv},
       eprint = {2307.10836},
 primaryClass = {astro-ph.SR},
       adsurl = {https://ui.adsabs.harvard.edu/abs/2023A&A...675A.180G}
}

@ARTICLE{LaresMartiz24,
       author = {{Lares-Martiz}, Mariel and {Buzasi}, Derek and {Oswalt}, Terry and {Confeiteiro}, Krystian and {Gee}, Ahnika and {Guida}, Luca and {Reynolds}, Ryan and {Walls}, Melinda},
        title = "{How Reliable are Rotation Period Determinations from TESS Data?}",
      journal = {Research Notes of the American Astronomical Society},
         year = 2024,
        month = may,
       volume = {8},
       number = {5},
          eid = {132},
        pages = {132},
          doi = {10.3847/2515-5172/ad4a7c},
       adsurl = {https://ui.adsabs.harvard.edu/abs/2024RNAAS...8..132L}
}

@ARTICLE{Barnes07,
       author = {{Barnes}, Sydney A.},
        title = "{Ages for Illustrative Field Stars Using Gyrochronology: Viability, Limitations, and Errors}",
      journal = {\apj},
         year = 2007,
        month = nov,
       volume = {669},
       number = {2},
        pages = {1167-1189},
          doi = {10.1086/519295},
archivePrefix = {arXiv},
       eprint = {0704.3068},
 primaryClass = {astro-ph},
       adsurl = {https://ui.adsabs.harvard.edu/abs/2007ApJ...669.1167B}
}

@ARTICLE{Mamajek08,
       author = {{Mamajek}, Eric E. and {Hillenbrand}, Lynne A.},
        title = "{Improved Age Estimation for Solar-Type Dwarfs Using Activity-Rotation Diagnostics}",
      journal = {\apj},
         year = 2008,
        month = nov,
       volume = {687},
       number = {2},
        pages = {1264-1293},
          doi = {10.1086/591785},
archivePrefix = {arXiv},
       eprint = {0807.1686},
 primaryClass = {astro-ph},
       adsurl = {https://ui.adsabs.harvard.edu/abs/2008ApJ...687.1264M}
}

@ARTICLE{SilvaBeyer23,
       author = {{Silva-Beyer}, Joaqu{\'\i}n and {Godoy-Rivera}, Diego and {Chanam{\'e}}, Julio},
        title = "{The breakdown of current gyrochronology as evidenced by old coeval stars}",
      journal = {\mnras},
         year = 2023,
        month = aug,
       volume = {523},
       number = {4},
        pages = {5947-5961},
          doi = {10.1093/mnras/stad1803},
archivePrefix = {arXiv},
       eprint = {2210.01137},
 primaryClass = {astro-ph.SR},
       adsurl = {https://ui.adsabs.harvard.edu/abs/2023MNRAS.523.5947S}
}

@ARTICLE{Deacon16,
       author = {{Deacon}, N.~R. and {Kraus}, A.~L. and {Mann}, A.~W. and {Magnier}, E.~A. and {Chambers}, K.~C. and {Wainscoat}, R.~J. and {Tonry}, J.~L. and {Kaiser}, N. and {Waters}, C. and {Flewelling}, H. and {Hodapp}, K.~W. and {Burgett}, W.~S.},
        title = "{A Pan-STARRS 1 study of the relationship between wide binarity and planet occurrence in the Kepler field}",
      journal = {\mnras},
         year = 2016,
        month = feb,
       volume = {455},
       number = {4},
        pages = {4212-4230},
          doi = {10.1093/mnras/stv2132},
archivePrefix = {arXiv},
       eprint = {1509.04712},
 primaryClass = {astro-ph.SR},
       adsurl = {https://ui.adsabs.harvard.edu/abs/2016MNRAS.455.4212D}
}

@ARTICLE{Otani22,
       author = {{Otani}, Tomomi and {von Hippel}, Ted and {Buzasi}, Derek and {Oswalt}, T.~D. and {Stone-Martinez}, Alexander and {Majewski}, Patrice},
        title = "{A Monte Carlo Method for Evaluating Empirical Gyrochronology Models and Its Application to Wide Binary Benchmarks}",
      journal = {\apj},
         year = 2022,
        month = may,
       volume = {930},
       number = {1},
          eid = {36},
        pages = {36},
          doi = {10.3847/1538-4357/ac6035},
archivePrefix = {arXiv},
       eprint = {2105.07266},
 primaryClass = {astro-ph.SR},
       adsurl = {https://ui.adsabs.harvard.edu/abs/2022ApJ...930...36O}
}

@ARTICLE{Andrews19,
       author = {{Andrews}, Jeff J. and {Anguiano}, Borja and {Chanam{\'e}}, Julio and {Ag{\"u}eros}, Marcel A. and {Lewis}, Hannah M. and {Hayes}, Christian R. and {Majewski}, Steven R.},
        title = "{Using APOGEE Wide Binaries to Test Chemical Tagging with Dwarf Stars}",
      journal = {\apj},
         year = 2019,
        month = jan,
       volume = {871},
       number = {1},
          eid = {42},
        pages = {42},
          doi = {10.3847/1538-4357/aaf502},
archivePrefix = {arXiv},
       eprint = {1811.12032},
 primaryClass = {astro-ph.SR},
       adsurl = {https://ui.adsabs.harvard.edu/abs/2019ApJ...871...42A}
}

@ARTICLE{Hollands24,
       author = {{Hollands}, M.~A. and {Littlefair}, S.~P. and {Parsons}, S.~G.},
        title = "{Measuring the initial-final mass relation using wide double white dwarf binaries from Gaia DR3}",
      journal = {\mnras},
         year = 2024,
        month = jan,
       volume = {527},
       number = {3},
        pages = {9061-9117},
          doi = {10.1093/mnras/stad3729},
archivePrefix = {arXiv},
       eprint = {2311.14801},
 primaryClass = {astro-ph.SR},
       adsurl = {https://ui.adsabs.harvard.edu/abs/2024MNRAS.527.9061H}
}

@ARTICLE{RojasAyala10,
       author = {{Rojas-Ayala}, B{\'a}rbara and {Covey}, Kevin R. and {Muirhead}, Philip S. and {Lloyd}, James P.},
        title = "{Metal-rich M-Dwarf Planet Hosts: Metallicities with K-band Spectra}",
      journal = {\apjl},
         year = 2010,
        month = sep,
       volume = {720},
       number = {1},
        pages = {L113-L118},
          doi = {10.1088/2041-8205/720/1/L113},
archivePrefix = {arXiv},
       eprint = {1007.4593},
 primaryClass = {astro-ph.SR},
       adsurl = {https://ui.adsabs.harvard.edu/abs/2010ApJ...720L.113R}
}

@ARTICLE{Montes18,
       author = {{Montes}, D. and {Gonz{\'a}lez-Peinado}, R. and {Tabernero}, H.~M. and {Caballero}, J.~A. and {Marfil}, E. and {Alonso-Floriano}, F.~J. and {Cort{\'e}s-Contreras}, M. and {Gonz{\'a}lez Hern{\'a}ndez}, J.~I. and {Klutsch}, A. and {Moreno-J{\'o}dar}, C.},
        title = "{Calibrating the metallicity of M dwarfs in wide physical binaries with F-, G-, and K-primaries - I: High-resolution spectroscopy with HERMES: stellar parameters, abundances, and kinematics}",
      journal = {\mnras},
         year = 2018,
        month = sep,
       volume = {479},
       number = {1},
        pages = {1332-1382},
          doi = {10.1093/mnras/sty1295},
archivePrefix = {arXiv},
       eprint = {1805.05394},
 primaryClass = {astro-ph.SR},
       adsurl = {https://ui.adsabs.harvard.edu/abs/2018MNRAS.479.1332M}
}

@ARTICLE{Garces11,
       author = {{Garc{\'e}s}, A. and {Catal{\'a}n}, S. and {Ribas}, I.},
        title = "{Time evolution of high-energy emissions of low-mass stars. I. Age determination using stellar chronology with white dwarfs in wide binaries}",
      journal = {\aap},
         year = 2011,
        month = jul,
       volume = {531},
          eid = {A7},
        pages = {A7},
          doi = {10.1051/0004-6361/201116775},
archivePrefix = {arXiv},
       eprint = {1105.0287},
 primaryClass = {astro-ph.SR},
       adsurl = {https://ui.adsabs.harvard.edu/abs/2011A&A...531A...7G}
}

@ARTICLE{Chaname12,
       author = {{Chanam{\'e}}, Julio and {Ram{\'\i}rez}, Iv{\'a}n},
        title = "{Toward Precise Ages for Single Stars in the Field. Gyrochronology Constraints at Several Gyr Using Wide Binaries. I. Ages for Initial Sample}",
      journal = {\apj},
         year = 2012,
        month = feb,
       volume = {746},
       number = {1},
          eid = {102},
        pages = {102},
          doi = {10.1088/0004-637X/746/1/102},
archivePrefix = {arXiv},
       eprint = {1109.0013},
 primaryClass = {astro-ph.SR},
       adsurl = {https://ui.adsabs.harvard.edu/abs/2012ApJ...746..102C}
}

@ARTICLE{Zhao12,
       author = {{Zhao}, J.~K. and {Oswalt}, T.~D. and {Willson}, L.~A. and {Wang}, Q. and {Zhao}, G.},
        title = "{The Initial-Final Mass Relation among White Dwarfs in Wide Binaries}",
      journal = {\apj},
         year = 2012,
        month = feb,
       volume = {746},
       number = {2},
          eid = {144},
        pages = {144},
          doi = {10.1088/0004-637X/746/2/144},
archivePrefix = {arXiv},
       eprint = {1112.0281},
 primaryClass = {astro-ph.SR},
       adsurl = {https://ui.adsabs.harvard.edu/abs/2012ApJ...746..144Z}
}

@ARTICLE{Andrews15,
       author = {{Andrews}, Jeff J. and {Ag{\"u}eros}, Marcel A. and {Gianninas}, A. and {Kilic}, Mukremin and {Dhital}, Saurav and {Anderson}, Scott F.},
        title = "{Constraints on the Initial-Final Mass Relation from Wide Double White Dwarfs}",
      journal = {\apj},
         year = 2015,
        month = dec,
       volume = {815},
       number = {1},
          eid = {63},
        pages = {63},
          doi = {10.1088/0004-637X/815/1/63},
archivePrefix = {arXiv},
       eprint = {1510.06107},
 primaryClass = {astro-ph.SR},
       adsurl = {https://ui.adsabs.harvard.edu/abs/2015ApJ...815...63A}
}

@ARTICLE{RebassaMansergas16,
       author = {{Rebassa-Mansergas}, A. and {Anguiano}, B. and {Garc{\'\i}a-Berro}, E. and {Freeman}, K.~C. and {Cojocaru}, R. and {Manser}, C.~J. and {Pala}, A.~F. and {G{\"a}nsicke}, B.~T. and {Liu}, X. -W.},
        title = "{The age-metallicity relation in the solar neighbourhood from a pilot sample of white dwarf-main sequence binaries}",
      journal = {\mnras},
         year = 2016,
        month = dec,
       volume = {463},
       number = {2},
        pages = {1137-1143},
          doi = {10.1093/mnras/stw2021},
archivePrefix = {arXiv},
       eprint = {1608.03064},
 primaryClass = {astro-ph.SR},
       adsurl = {https://ui.adsabs.harvard.edu/abs/2016MNRAS.463.1137R}
}

@ARTICLE{Makarov08,
       author = {{Makarov}, V.~V. and {Zacharias}, N. and {Hennessy}, G.~S.},
        title = "{Common Proper Motion Companions to Nearby Stars: Ages and Evolution}",
      journal = {\apj},
         year = 2008,
        month = nov,
       volume = {687},
       number = {1},
        pages = {566-578},
          doi = {10.1086/591638},
archivePrefix = {arXiv},
       eprint = {0808.3414},
 primaryClass = {astro-ph},
       adsurl = {https://ui.adsabs.harvard.edu/abs/2008ApJ...687..566M}
}

@ARTICLE{Kraus09,
       author = {{Kraus}, Adam L. and {Hillenbrand}, Lynne A.},
        title = "{The Coevality of Young Binary Systems}",
      journal = {\apj},
         year = 2009,
        month = oct,
       volume = {704},
       number = {1},
        pages = {531-547},
          doi = {10.1088/0004-637X/704/1/531},
archivePrefix = {arXiv},
       eprint = {0909.0509},
 primaryClass = {astro-ph.SR},
       adsurl = {https://ui.adsabs.harvard.edu/abs/2009ApJ...704..531K}
}

@ARTICLE{Andrae23,
       author = {{Andrae}, Ren{\'e} and {Rix}, Hans-Walter and {Chandra}, Vedant},
        title = "{Robust Data-driven Metallicities for 175 Million Stars from Gaia XP Spectra}",
      journal = {\apjs},
         year = 2023,
        month = jul,
       volume = {267},
       number = {1},
          eid = {8},
        pages = {8},
          doi = {10.3847/1538-4365/acd53e},
archivePrefix = {arXiv},
       eprint = {2302.02611},
 primaryClass = {astro-ph.SR},
       adsurl = {https://ui.adsabs.harvard.edu/abs/2023ApJS..267....8A}
}

@ARTICLE{Deason24,
       author = {{Deason}, Alis J. and {Belokurov}, Vasily},
        title = "{Galactic Archaeology with Gaia}",
      journal = {\nar},
         year = 2024,
        month = dec,
       volume = {99},
          eid = {101706},
        pages = {101706},
          doi = {10.1016/j.newar.2024.101706},
archivePrefix = {arXiv},
       eprint = {2402.12443},
 primaryClass = {astro-ph.GA},
       adsurl = {https://ui.adsabs.harvard.edu/abs/2024NewAR..9901706D}
}

@ARTICLE{Sanders18,
       author = {{Sanders}, Jason L. and {Das}, Payel},
        title = "{Isochrone ages for {\ensuremath{\sim}}3 million stars with the second Gaia data release}",
      journal = {\mnras},
         year = 2018,
        month = dec,
       volume = {481},
       number = {3},
        pages = {4093-4110},
          doi = {10.1093/mnras/sty2490},
archivePrefix = {arXiv},
       eprint = {1806.02324},
 primaryClass = {astro-ph.GA},
       adsurl = {https://ui.adsabs.harvard.edu/abs/2018MNRAS.481.4093S}
}

@ARTICLE{Bonfils05,
       author = {{Bonfils}, X. and {Delfosse}, X. and {Udry}, S. and {Santos}, N.~C. and {Forveille}, T. and {S{\'e}gransan}, D.},
        title = "{Metallicity of M dwarfs. I. A photometric calibration and the impact on the mass-luminosity relation at the bottom of the main sequence}",
      journal = {\aap},
         year = 2005,
        month = nov,
       volume = {442},
       number = {2},
        pages = {635-642},
          doi = {10.1051/0004-6361:20053046},
archivePrefix = {arXiv},
       eprint = {astro-ph/0503260},
 primaryClass = {astro-ph},
       adsurl = {https://ui.adsabs.harvard.edu/abs/2005A&A...442..635B}
}

@ARTICLE{Johnson09,
       author = {{Johnson}, John Asher and {Apps}, Kevin},
        title = "{On the Metal Richness of M Dwarfs with Planets}",
      journal = {\apj},
         year = 2009,
        month = jul,
       volume = {699},
       number = {2},
        pages = {933-937},
          doi = {10.1088/0004-637X/699/2/933},
archivePrefix = {arXiv},
       eprint = {0904.3092},
 primaryClass = {astro-ph.EP},
       adsurl = {https://ui.adsabs.harvard.edu/abs/2009ApJ...699..933J}
}

@ARTICLE{Helmi20,
       author = {{Helmi}, Amina},
        title = "{Streams, Substructures, and the Early History of the Milky Way}",
      journal = {\araa},
         year = 2020,
        month = aug,
       volume = {58},
        pages = {205-256},
          doi = {10.1146/annurev-astro-032620-021917},
archivePrefix = {arXiv},
       eprint = {2002.04340},
 primaryClass = {astro-ph.GA},
       adsurl = {https://ui.adsabs.harvard.edu/abs/2020ARA&A..58..205H}
}

@ARTICLE{Queiroz18,
       author = {{Queiroz}, A.~B.~A. and {Anders}, F. and {Santiago}, B.~X. and {Chiappini}, C. and {Steinmetz}, M. and {Dal Ponte}, M. and {Stassun}, K.~G. and {da Costa}, L.~N. and {Maia}, M.~A.~G. and {Crestani}, J. and {Beers}, T.~C. and {Fern{\'a}ndez-Trincado}, J.~G. and {Garc{\'\i}a-Hern{\'a}ndez}, D.~A. and {Roman-Lopes}, A. and {Zamora}, O.},
        title = "{StarHorse: a Bayesian tool for determining stellar masses, ages, distances, and extinctions for field stars}",
      journal = {\mnras},
         year = 2018,
        month = may,
       volume = {476},
       number = {2},
        pages = {2556-2583},
          doi = {10.1093/mnras/sty330},
archivePrefix = {arXiv},
       eprint = {1710.09970},
 primaryClass = {astro-ph.IM},
       adsurl = {https://ui.adsabs.harvard.edu/abs/2018MNRAS.476.2556Q}
}

@ARTICLE{Conroy22,
       author = {{Conroy}, Charlie and {Weinberg}, David H. and {Naidu}, Rohan P. and {Buck}, Tobias and {Johnson}, James W. and {Cargile}, Phillip and {Bonaca}, Ana and {Caldwell}, Nelson and {Chandra}, Vedant and {Han}, Jiwon Jesse and {Johnson}, Benjamin D. and {Speagle}, Joshua S. and {Ting}, Yuan-Sen and {Woody}, Turner and {Zaritsky}, Dennis},
        title = "{Birth of the Galactic Disk Revealed by the H3 Survey}",
      journal = {arXiv e-prints},
         year = 2022,
        month = apr,
          eid = {arXiv:2204.02989},
        pages = {arXiv:2204.02989},
          doi = {10.48550/arXiv.2204.02989},
archivePrefix = {arXiv},
       eprint = {2204.02989},
 primaryClass = {astro-ph.GA},
       adsurl = {https://ui.adsabs.harvard.edu/abs/2022arXiv220402989C}
}

@ARTICLE{Hawkins20,
       author = {{Hawkins}, Keith and {Lucey}, Madeline and {Ting}, Yuan-Sen and {Ji}, Alexander and {Katzberg}, Dustin and {Thompson}, Megan and {El-Badry}, Kareem and {Teske}, Johanna and {Nelson}, Tyler and {Carrillo}, Andreia},
        title = "{Identical or fraternal twins? The chemical homogeneity of wide binaries from Gaia DR2}",
      journal = {\mnras},
         year = 2020,
        month = feb,
       volume = {492},
       number = {1},
        pages = {1164-1179},
          doi = {10.1093/mnras/stz3132},
archivePrefix = {arXiv},
       eprint = {1912.08895},
 primaryClass = {astro-ph.SR},
       adsurl = {https://ui.adsabs.harvard.edu/abs/2020MNRAS.492.1164H}
}

@ARTICLE{Mann13,
       author = {{Mann}, Andrew W. and {Brewer}, John M. and {Gaidos}, Eric and {L{\'e}pine}, S{\'e}bastien and {Hilton}, Eric J.},
        title = "{Prospecting in Late-type Dwarfs: A Calibration of Infrared and Visible Spectroscopic Metallicities of Late K and M Dwarfs Spanning 1.5 dex}",
      journal = {\aj},
         year = 2013,
        month = feb,
       volume = {145},
       number = {2},
          eid = {52},
        pages = {52},
          doi = {10.1088/0004-6256/145/2/52},
archivePrefix = {arXiv},
       eprint = {1211.4630},
 primaryClass = {astro-ph.SR},
       adsurl = {https://ui.adsabs.harvard.edu/abs/2013AJ....145...52M}
}

@ARTICLE{Masseron15,
       author = {{Masseron}, T. and {Gilmore}, G.},
        title = "{Carbon, nitrogen and {\ensuremath{\alpha}}-element abundances determine the formation sequence of the Galactic thick and thin discs}",
      journal = {\mnras},
         year = 2015,
        month = oct,
       volume = {453},
       number = {2},
        pages = {1855-1866},
          doi = {10.1093/mnras/stv1731},
archivePrefix = {arXiv},
       eprint = {1503.00537},
 primaryClass = {astro-ph.SR},
       adsurl = {https://ui.adsabs.harvard.edu/abs/2015MNRAS.453.1855M}
}

@ARTICLE{Martig16,
       author = {{Martig}, Marie and {Fouesneau}, Morgan and {Rix}, Hans-Walter and {Ness}, Melissa and {M{\'e}sz{\'a}ros}, Szabolcs and {Garc{\'\i}a-Hern{\'a}ndez}, D.~A. and {Pinsonneault}, Marc and {Serenelli}, Aldo and {Silva Aguirre}, Victor and {Zamora}, Olga},
        title = "{Red giant masses and ages derived from carbon and nitrogen abundances}",
      journal = {\mnras},
         year = 2016,
        month = mar,
       volume = {456},
       number = {4},
        pages = {3655-3670},
          doi = {10.1093/mnras/stv2830},
archivePrefix = {arXiv},
       eprint = {1511.08203},
 primaryClass = {astro-ph.SR},
       adsurl = {https://ui.adsabs.harvard.edu/abs/2016MNRAS.456.3655M}
}

@ARTICLE{Nissen15,
       author = {{Nissen}, P.~E.},
        title = "{High-precision abundances of elements in solar twin stars. Trends with stellar age and elemental condensation temperature}",
      journal = {\aap},
         year = 2015,
        month = jul,
       volume = {579},
          eid = {A52},
        pages = {A52},
          doi = {10.1051/0004-6361/201526269},
archivePrefix = {arXiv},
       eprint = {1504.07598},
 primaryClass = {astro-ph.SR},
       adsurl = {https://ui.adsabs.harvard.edu/abs/2015A&A...579A..52N}
}

@BOOK{Pagel09,
       author = {{Pagel}, Bernard E.~J.},
        title = "{Nucleosynthesis and Chemical Evolution of Galaxies}",
         year = 2009,
       adsurl = {https://ui.adsabs.harvard.edu/abs/2009nceg.book.....P}
}
\bibliographystyle{mnras}

\end{document}